\documentstyle[12pt]{article}
\newcommand{\beq}{\begin{equation}}
\newcommand{\eeq}{\end{equation}}
\newcommand{\bea}{\begin{eqnarray}}
\newcommand{\eea}{\end{eqnarray}}

\begin{document}
\setcounter{page}{0}
\topmargin 0pt
\oddsidemargin 5mm
\renewcommand{\thefootnote}{\fnsymbol{footnote}}
\newpage
\setcounter{page}{0}
\begin{titlepage}

\begin{flushright}
QMW-PH-95-46\\
{\bf hep-th/9512014}\\
November $23$, $1995$
\end{flushright}
\vspace{0.5cm}
\begin{center}
{\Large {\bf Is a truly marginal perturbation of the $G_k\times G_k$
WZNW model
at $k=-2c_V(G)$ an exception to the rule?}} \\
\vspace{1.8cm}
\vspace{0.5cm}
{Oleg A.
Soloviev
\footnote{e-mail: O.A.Soloviev@QMW.AC.UK}}\\
\vspace{0.5cm}
{\em Physics Department, Queen Mary and Westfield College, \\
Mile End Road, London E1 4NS, United Kingdom}\\
\vspace{0.5cm}
\renewcommand{\thefootnote}{\arabic{footnote}}
\setcounter{footnote}{0}
\begin{abstract}
{It is shown that there exists a truly marginal deformation of the
direct sum
of two $G_k$ WZNW models at $k=-2c_V(G)$ (where $c_V(G)$ is the
eigenvalue  of
the quadratic Casimir operator in the adjoint representation of the
group $G$)
which does not seem to fit the Chaudhuri-Schwartz criterion for truly
marginal
perturbations. In addition, a continuous family of WZNW models is
constructed.}
\end{abstract}
\vspace{0.5cm}
\centerline{November 1995}
 \end{center}
\end{titlepage}
\newpage
\section{Introduction}
The sigma-model interpretation of conformal field theories plays a
crucial role
in bridging string theory with point-like physics.  The latter is
understood as
the low energy limit of the former. Crossing over this bridge leads
one to
discover quite fascinating features of the stringy description of
space-time
and forces acting in it. One of these features is that two
space-times endowed
with entirely different geometries and topologies could, in fact, be
completely
undistinguishable as they correspond to one and the same CFT. In
string theory
these two target spaces are described by two different conformal
sigma-models
which are related to each other via transformations of some coupling
constants
which range in a continuous interval. If at each point in this
interval the
conformal symmetry is preserved, then the two sigma models are said
to
correspond to one and the same CFT. From this point of view, the
study of
conformal sigma models admitting such continuous transformations is
important
for a better understanding string theory predictions.

The issue of continuous conformal deformations of sigma-models has
attracted
much of attention in the recent years (see e.g.
\cite{Kadanoff}-\cite{Evans}).
Some of these deformations are understood as integrated
infinitesimally small
perturbations on a given CFT. For these perturbations to be
integrable, they
have to be generated by truly marginal operators, which are primary
conformal
operators with dimension (1,1) all the way along the perturbation.
This in turn
requires the renormalization group beta functions of the
corresponding
perturbation parameters to vanish at all values of these parameters.

Given a CFT, one can define a perturbed CFT as follows
\begin{equation}
S(\epsilon)=S^{*}-\epsilon\int d^2z~O(z,\bar z), \end{equation}
where $O$ is a perturbation operator and $\epsilon$ is some small
parameter
(henceforth we will be discussing only one-parameter deformations,
which impose
some renormalizability conditions which will be supposed to be
fulfilled). The
action (1.1) describes a one-parameter perturbation of a CFT given by
$S^{*}$.
In all cases, the operator $O(z,\bar z)$ can be presented in the
following
factorized form
\begin{equation}
O(z,\bar z)=\sum_{A,\bar A}\lambda_{A\bar A}J^A(z)\bar J^{\bar
A}(\bar
z),\end{equation}
where $J^A,~\bar J^{\bar A}$ are some analytic functions of $z$ and
$\bar z$
respectively, whereas $\lambda_{A\bar A}$ are some constant
coefficients.
$J^A,~\bar J^{\bar A}$ can be called currents. However, it is
unnecessary that
they form any current algebra. In the case when $J^A,~\bar J^{\bar
A}$ do form
a certain algebra, there has been suggested a criterion for the
perturbation
(1.1) to be truly marginal \cite{Chaudhuri}. The criterion is
formulated as
follows: a perturbation of the type (1.2) is truly (or integrably)
marginal if
and only if the coefficients $\lambda_{A\bar A}$ are non-zero only
for $A,~\bar
A$ in some abelian subalgebra of the full current algebra. Stated as
above the
criterion seems to provide us with both necessary and sufficient
conditions.
There have been some examples examined which perfectly fit into the
given
criterion \cite{Chaudhuri}.

The aim of the present paper is to exhibit one particular example of
a truly
marginal perturbation which does not obey the Chaudhuri-Schwartz
criterion. We
shall show that the direct sum of two $G_k$ WZNW models at the
special value of
the level $k=-2c_V(G)$ admits a continuous perturbation by a truly
marginal
operator which cannot be presented as a product of abelian currents.
In this
way we shall prove that the CS criterion is sufficient but not
necessary.

The paper is organized as follows. In section 2 we consider the
action of a
direct sum of two $G_k$ WZNW models perturbed by the Thirring like
current-current interaction. We show that the perturbation of the sum
of two
WZNW models can be recast into a perturbation of one WZNW model which
appears
to be quite handy for the further analysis of conformal symmetry. In
section 3
we demonstrate that the perturbation operator becomes a singular
vector when
$k=-2c_V$. Moreover, it forms a closed fusion algebra. In section 4
we study
the effect of the singular perturbation on the conformal symmetry. It
turns out
that the given perturbation does not break the conformal invariance.
In section
5 we introduce a continuous family of WZNW models with the
Wess-Zumino coupling
constant $k=-2c_V$ and the sigma-model coupling constant being
arbitrary.
Section 6 contains our conclusion.

\section{$G\times G$ WZNW model and its perturbation}

By the $G\times G$ WZNW model we understand the direct product of two
identical
$G$ WZNW models with level $k$. The action is given by
\begin{equation}
S_{G\times G}=S_{WZNW}(g_1,k)~+~S_{WZNW}(g_2,k),\end{equation}
where
\begin{equation}
S_{WZNW}(g_{1,2},k)=-{k\over4\pi}\left\{\int
x{Tr}|g^{-1}_{1,2}\mbox{d}g_{1,2}|^2+{i\over3}\int
\mbox{d}^{-1}\mbox{Tr}(g^{-1}_{1,2}\mbox{d}g_{1,2})^3\right\},
\end{equation}
with $g_1$ and $g_2$ taking values in the Lie group $G$.

The theory in eq. (2.3) can be generalized in a way which does not
lead to
missing the underlying affine symmetries. Namely, one can add to the
sum (2.3)
the following interaction term \cite{Soloviev}
\begin{equation}
S_I=-{k^2\over\pi}\int d^2z\mbox{Tr}^2(g^{-1}_1\partial
g_1~S~\bar\partial
g_2g^{-1}_2),\end{equation}
with the coupling $S$ belonging to the direct product of two Lie
algebras
${\cal G}\times{\cal G}$,
\begin{equation}
S=S_{ab}~t^a\otimes t^b,\end{equation}
where
\begin{equation}
[t^a,t^b]=f^{ab}_ct^c,\end{equation}
with $f^{ab}_c$ the structure constants of ${\cal G}$.

It is convenient to present the coupling matrix $S_{ab}$ in the
following form
\begin{equation}
S_{ab}=\sigma\hat S_{ab},\end{equation}
where $\sigma$ is a constant and $\hat S_{ab}$ is another matrix
which will be
thought of as being fixed. Then for small $\sigma$, the interaction
term (2.5)
can be observed as a perturbation of the type (1.2) on the direct sum
of two
WZNW models. Obviously, in this case, the currents $J^A,~\bar J^{\bar
A}$ are
identified with the affine currents and as such form affine algebras:
\begin{eqnarray}
J^A&\longrightarrow &J^a_1=-{k\over2}\mbox{Tr}(g^{-1}_1\partial
g_1t^a),\nonumber\\ & & \\
\bar J^{\bar A}&\longrightarrow &\bar J^a_2=-{k\over2}
\mbox{Tr}(\bar\partial g_2g^{-1}_2t^a).\nonumber\end{eqnarray}
Correspondingly, the perturbation operator $O$ is
\begin{equation}
O=\hat S_{ab}~J^a_1\bar J^b_2,\end{equation}
and the perturbation parameter is given by
\begin{equation}
\epsilon={4\sigma\over\pi}.\end{equation}

Clearly, the operator $O$ is a marginal primary operator in the
$G\times G$
WZNW model. However, in the perturbed theory it may acquire anomalous
conformal
dimension. If it is the case, then the conformal symmetry gets broken
under the
perturbation. In \cite{Chaudhuri}, the authors calculated the
two-point
function of the perturbation operator in order to find the change in
the
conformal dimension of $O$. Their explicit computations went as far
as to terms
of order $\epsilon^2$. We think that the analysis in \cite{Chaudhuri}
is
incomplete and some interesting cases were missing. We are going to
show that
there is a more effective method of calculating perturbative effects
which will
allow us to prove that the results of \cite{Chaudhuri} give rise to
sufficient
but not necessary conditions for $O$ to be truly marginal.

We would like to start with recasting the $G\times G$ WZNW theory
perturbed by
the operator $O$ into somewhat different form. Let us make the
following change
of variables \cite{Hull}
\begin{eqnarray}
g_1&\longrightarrow& \tilde g_1,\nonumber\\ & & \\
g_2&\longrightarrow& h(\tilde g_1)\cdot\tilde
g_2,\nonumber\end{eqnarray}
where $\tilde g_1,~\tilde g_2$ are new variables, whereas the
function
$h(\tilde g_1)$ is the solution of the following equation
\begin{equation}
\partial hh^{-1}=-2k\mbox{Tr}S\tilde g_1^{-1}
\partial\tilde g_1.\end{equation}

In terms of the new variables, the action takes the form
\cite{Soloviev},\cite{Hull}
\begin{eqnarray}
\tilde S_{G\times G}&=&S_{WZNW}(\tilde g_2,k)~+~S_{WZNW}(\tilde
g_1,k)\nonumber\\ & & \\
&+&{k^3\over\pi}\int d^2z\mbox{Tr}\left(\mbox{Tr}S
\tilde g_1^{-1}\partial\tilde g_1~\mbox{Tr}S\tilde g_1^{-1}
\bar\partial\tilde g_1\right)~+~{\cal O}(S^3).\nonumber\end{eqnarray}
The important point to be made is that after this change of
variables, the
field $\tilde g_2$ completely decouples from $\tilde g_1$. As one can
see
$\tilde g_2$ is governed simply by a WZNW action, whereas the action
for
$\tilde g_1$ is more complicated. Fortunately, this action can be
understood as
a perturbed WZNW model \cite{Soloviev},\cite{Hull}
\begin{equation}
S(\tilde g_1)=S_{WZNW}(\tilde g_1,k)~-~
\tilde\epsilon\int d^2z~\tilde O(z,\bar z),\end{equation}
where
\begin{equation}
\tilde O(z,\bar z)=\hat S_{ac}\hat S_{bc}~\tilde J^a\tilde{\bar
J^{\bar
b}}\tilde\phi^{b\bar b}.\end{equation}
Here
\begin{eqnarray}
\tilde J&=&-{k\over2}\tilde g^{-1}_1\partial\tilde g_1,\nonumber\\
\tilde{\bar J}&=&-{k\over2}\bar\partial
\tilde g_1\tilde g_1^{-1},\nonumber\\& &\\
\tilde\phi^{a\bar a}&=&\mbox{Tr}(\tilde g_1^{-1}t^a\tilde g_1t^{\bar
a}),\nonumber\\
\tilde\epsilon&=&{4\sigma^2\over\pi}.\nonumber\end{eqnarray}
Note that $\tilde O$ can be presented in the form (1.2), however, new
currents
will not form an affine algebra. Therefore, one cannot apply the CS
criterion
to the given operator.

In principle, one can further expand the action (2.14) in the
coupling
$\sigma$. However, higher order expansion is going to be much more
involved as
it requires a proper handling of non-local terms. In this paper we
will
restrict ourselves to the approximation given by eq. (2.15) which is
local.

\section{$k=-2c_V(G)$}

In order for the perturbed theory (2.15) to be a well defined quantum
field
theory, the perturbation $\tilde O$ has to obey some consistency
conditions.
The crucial one is the renormalizability condition, which amounts to
a
condition on the OPE of the operator $\tilde O$ with itself
\cite{Hull}. This
in turn imposes restrictions on the matrix $\hat S_{ab}$. There are a
number of
matrices which satisfy the given consistency condition. We will be
interested
in the matrix $\hat S_{ab}$ given by
\begin{equation}
\hat S_{ab}=\delta_{ab}.\end{equation}
It is not difficult to see that $\tilde O$ is a Virasoro primary
operator with
the conformal dimension
\begin{equation}
\Delta=1+{c_V(G)\over k+c_V(G)},\end{equation}
where the quantity $c_V(G)$ is defined as follows
\begin{equation}
f^{ac}_df^{bd}_c=-c_V~\delta^{ab}.\end{equation}
Thus, for all positive $k$, the operator $\tilde O$ is an irrelevant
conformal
operator from the point of view of the renormalization procedure.
This
circumstance would cause the infra-red divergences in the course of
quantization of the perturbed theory (2.15). Therefore in general,
one has to
put the system into a finite box in order to avoid the IR troubles.
However,
there is one particular value of $k$ at which dramatic
simplifications occur.

Let us take
\begin{equation}
k=-2c_V,\end{equation}
where $c_V$ is given by eq. (3.20). In this case,
\begin{equation}
\Delta=0\end{equation}
for all $G$. Let us now compute the norm of the given operator.
\begin{eqnarray}
||\tilde O||^2&=&\langle0|\tilde O^\dagger(0)\tilde O(0)|0\rangle
=\langle\tilde \phi^{a\bar a}|\tilde J^a_1\tilde{\bar J^{\bar
a}_1}\tilde
J^b_{-1}\tilde{\bar J^{\bar b}}_{-1}|\tilde\phi^{b\bar b}
\rangle\nonumber\\ & & \\
&=&(c_V+{k\over2})^2\langle\tilde\phi^{a\bar a}|\tilde\phi^{a\bar
a}\rangle.\nonumber\end{eqnarray}
We have used the standard conjugation rule of affine generators
\begin{equation}
\tilde J^\dagger_n=\tilde J_{-n},~~~~~~\tilde{\bar J~}^\dagger_n=
\tilde{\bar J_{-n}},\end{equation}
where
\begin{equation}
\tilde J_n=\oint{dz\over2\pi i}~z^n\tilde J(z),~~~~~~~\tilde{\bar
J_n}=\oint{d\bar z\over2\pi i}~\bar z^n\tilde{\bar J(}\bar z).
\end{equation}
As one can see at $k=-2c_V$, the norm given by eq. (3.23) vanishes,
\begin{equation}
||\tilde O||^2_{k=-2c_V}=0.\end{equation}
In other words, at the special value of $k$, the operator $\tilde O$
becomes a
singular vector in the spectrum of the $G$-WZNW model.

It is easy to show that eq. (3.26) is, in fact, a consequence of the
following
equality
\begin{equation}
\tilde J_1^a|\tilde O\rangle=0,\end{equation}
which implies
\begin{equation}
\tilde J^a_{n>1}|\tilde O\rangle=0.\end{equation}
Thus, as it is usually the case for singular vectors, $\tilde O$ is
both a
Virasoro primary and an affine primary. (Remember, that for
$k\ne2c_V$,
$\tilde O$ is just a Virasoro primary.) Also one can prove that
$\tilde O$
forms a closed algebra which results in the following relation
\begin{equation}
\tilde O(z,\bar z)~|\tilde O\rangle\sim|\tilde
O\rangle.\end{equation}
Indeed, one can check that \cite{Soloviev2}
\begin{equation}
\tilde J^a_1\left(\tilde O(z,\bar z)|\tilde O\rangle\right)
=0,\end{equation}
and
\begin{equation}
\tilde J^a_0\left(\tilde O(z,\bar z)|\tilde
O\rangle\right)=0.\end{equation}
Hence, the right hand side of eq. (3.29) must be an affine singular
primary
vector which is also a singlet with respect to the global group $G$.
The only
possibility for these conditions to be satisfied is given by eq.
(3.29).

\section{Perturbation by the singular vector}

Now we want to discuss the effect of the singular perturbation on the
conformal
symmetry of the perturbed theory. The latter is described by the
following
action
\begin{equation}
S(\epsilon)=S_{WZNW}(g,k)~-~\epsilon\int d^2z~O(z,\bar z).
\end{equation}
Here the operator $O$ coincides with the operator $\tilde O$. From
now on we
will omit tilde in all expressions. The important point to be made is
that the
given theory is renormalizable. This is due to the property (3.29).
The
renormalizability implies that the trace of the energy-momentum
tensor is
expressed as follows
\begin{equation}
\Theta=\beta(\epsilon)~O,\end{equation}
where $\beta$ is the renormalization group beta function,
\begin{equation}
\beta(\epsilon)=\mbox{d}\epsilon/\mbox{d}t.\end{equation}

We have shown in the previous section that $O$ ($\equiv\tilde O$) is
a singular
vector. Because of eq. (4.33), the trace $\Theta$ is a singular
vector as well.
Hence,  $\Theta$ must have no effect on the conformal symmetry of the
perturbed
theory. Let us show that the Virasoro central charge indeed does not
change.

According to Zamolodchikov's $c$-theorem \cite{Zamolodchikov}
\begin{equation}
{\mbox{d}c\over\mbox{d}t}=-12\beta^2\langle0|O(1)O(0)|0\rangle.
\end{equation}
There holds the following relation
\begin{equation}
O(0)|0\rangle=J^a_{-1}\bar J^{\bar a}_{-1}|\phi^{a\bar a}\rangle.
\end{equation}
Moreover, one can prove that
\begin{equation}
[J^a_{-1},O(z,\bar z)]=0.\end{equation}
Then commuting $J^a_{-1}$ in the correlator in eq. (4.35) to the
left, one
obtains
\begin{equation}
\langle0|O(1)O(0)|0\rangle=0\end{equation}
and, hence,
\begin{equation}
{\mbox{d}c\over\mbox{d}t}=0.\end{equation}
Thus, $c$ remains constant along the renormalization group flow
associated with
the singular perturbation, even though the beta function $\beta$ is
nonzero.
All in all, we arrive at a conclusion that the conformal symmetry is
not
affected by the singular perturbation.

Now we can return to the system of two interacting WZNW models which
we
introduced in section 2. Obviously, since the latter system is
related to the
perturbation we have just discussed above, the conformal symmetry of
the
interacting WZNW models does not get broken under a continuous
variation of the
coupling constant $\sigma$. In other words, the operator in eq.
(2.10) is truly
marginal. However, the parameter $\sigma$ may change only in a finite
interval.
The restriction comes from the fact that at $\sigma={1\over4c_V}$,
the system
of two interacting WZNW models acquires a gauge symmetry as one can
easily see
by using the Polyakov-Wiegmann formula \cite{Polyakov}. Thus, if one
moves
$\sigma$ in the negative direction, one can continuously proceed to
the value
${1\over4c_V}$. Whereas in the positive direction there do not seem
to occur
any restrictions. It is also possible that there may be a certain
duality
symmetry  which can mirror the restriction on the negative direction
into a
restriction on positive values of $\sigma$.

Now we come to our main conclusion that despite the fact that our
perturbation
operator does not satisfy the Chaudhuri-Schwartz criterion,
nevertheless, it is
a truly marginal operator. Of course, there is the issue of unitarity
of the
WZNW model at negative level $k$. However, we think that all negative
normed
states can be projected out by a BRST-like procedure.

\section{A continuous family of WZNW models at $k=-2c_V$}

In the previous sections we have exhibited that the perturbed WZNW
model (4.32)
arises as a lower order effective theory of two interacting WZNW
models. At the
same time one can consider the theory described by the action (4.32)
as a
fundamental quantum field theory, not as an effective model. In this
case, one
does not need to worry about higher order corrections in $\epsilon$,
because
the action (4.32) will be the whole theory. We are going to show that
this
theory is, in fact, very curious.

First of all, we want to show that the perturbation $O$ does not
break the
affine symmetry of the WZNW model at $k=-2c_V$. The affine symmetry
is
generated by the affine current $J^a(z)$. Thus, our aim is to compute
the
following commutator $[J^a(y),O(z,\bar z)]$, which can be understood
as equal
time commutator. To this end, we present $O(z,\bar z)$ in the
factorized form
\begin{equation}
O(z,\bar z)=O(z)\cdot\bar O(\bar z),\end{equation}
where
\begin{equation}
O(z)=:J^a(z)\phi^a(z):.\end{equation}
Here $\phi^a(z)$ is defined as follows
\begin{equation}
\phi^{a\bar a}(z,\bar z)=\phi^a(z)\cdot\bar\phi^{\bar a}(\bar z).
\end{equation}
Note that in general the splitting into holomorphic and
antiholomorphic parts
can be more tricky. Namely, one can write
\begin{equation}
\phi^{a\bar a}(z,\bar z)=\phi^a_i(z)\cdot K^{ij}\cdot
\bar\phi^{\bar a}_j(\bar z),\end{equation}
where $K^{ij}$ is some matrix which does not depend on $z,~\bar z$.
But for our
purposes, one can take $K^{ij}=1$.

We start with a definition of normal ordering in eq. (5.41). We
define it
according to
\begin{equation}
:J^a(z)\phi^a(z):\equiv\oint{d\zeta\over2\pi
i}~{J^a(\zeta)\phi^a(z)\over\zeta-z}.\end{equation}

Now let us compute
\begin{eqnarray}
[J^a(y),O(z)]&=&\oint{d\zeta\over2\pi
i}{1\over\zeta-z}\left\{[J^a(y),J^b(\zeta)]\phi^b(z)+
J^b(\zeta)[J^a(y),\phi^b(z)]\right\}\nonumber\\
&=&\oint{d\zeta\over2\pi i}{1\over\zeta-z}\left\{f^{ab}_cJ^c(\zeta)
\phi^b(z)\delta(y,\zeta)+{k\over2}\phi^a(z)\delta'(y,\zeta)+
f^{ab}_cJ^b(\zeta)\phi^c(z)\delta(y,z)\right\}\nonumber\\
&=&\oint{d\zeta\over2\pi i}{1\over\zeta-z}
\left\{{f^{ab}_cf^{cb}_d\over\zeta-z}
\phi^d(z)\delta(y,\zeta)+{k\over2}\phi^a(z)\delta'(y,\zeta)+
{f^{ab}_cf^{bc}_d\over\zeta-z}\phi^d(z)\delta(y,z)\right\}
\nonumber\\
&+&\oint{d\zeta\over2\pi i}{1\over\zeta-z}[f^{ab}_c
\Psi^{cb}(z)\delta(y,\zeta)+
f^{ac}_b\Psi^{cb}(z)
\delta(y,z)].\end{eqnarray}
We have used the following relations
\begin{eqnarray}
\left[J^a(y),J^b(z)\right]&=&f^{ab}_cJ^c(z)\delta(y,z)+
{k\over2}\delta^{ab}\delta'(y,z),\nonumber\\
\left[J^a(y),\phi^b(z)\right]&=&f^{ab}_c\phi^c(z)\delta(y,z),\\
\Psi^{cb}(z)&\equiv&:J^c(z)\phi^b(z):.\nonumber\end{eqnarray}
By taking contour integrals in formula (5.45), we obtain
\begin{equation}
[J^a(y),O(z)]=({k\over2}+c_V)\phi^a(z)\delta'(y,z).\end{equation}
Therefore, at $k=-2c_V$,
\begin{equation}
[J^a(y),O(z)]=0.\end{equation}
Correspondingly,
\begin{equation}
[J^a(y),O(z,\bar z)]=[\bar J^{\bar a}(\bar y),O(z,\bar z)]=0.
\end{equation}

Thus, the theory given by eq. (4.32) possesses both the conformal
symmetry and
the affine symmetry of the original WZNW model at $k=-2c_V$.

It is interesting to look at the classical limit of the operator
$O(z,\bar z)$. We find
\begin{equation}
O(z,\bar z)\longrightarrow-c_V^2\mbox{Tr}(\partial g\bar\partial
g^{-1}).\end{equation}
Hence, by perturbing the WZNW model at $k=-2c_V$ by the operator
(2.16), we
change effectively the sigma-model coupling constant. In other words,
the
classical theory is described by the following action
\begin{equation}
S(\lambda)={1\over4\lambda}\int d^2x~
\mbox{Tr}(\partial_\mu g\partial^\mu g^{-1})~-~2c_V\Gamma,
\end{equation}
where $\Gamma$ is the Wess-Zumino term. The above analysis suggests
that the
given theory is conformal for arbitrary (negative) coupling constant
$\lambda$.
This theory describes a continuous family of WZNW models with the
special
Wess-Zumino coupling constant $k=-2c_V$.

\section{Conclusion}

We have exhibited one example of a truly marginal perturbation which
does not
satisfy the Chaudhuri-Schwartz criterion. By using this new
perturbation, we
have found a continuous family of WZNW models with arbitrary
sigma-model
coupling constant.

I would like to thank Henric Rhedin and Steve Thomas for useful
discussions and
the PPARC for financial support.

\end{document}